\documentstyle[12pt]{article}
\begin{document}

\begin{center}
{\Large \bf
{ ANTIGRAVITATING BUBBLES }
}\\
\vskip 1.0cm
{ Andro BARNAVELI and Merab GOGBERASHVILI } \\
\vskip 0.5cm
{\it
 {Institute of Physics of the Georgian Academy of Sciences,}\\
 {Tamarashvili str. 6, Tbilisi 380077, Republic of Georgia.}\\
 {(E-mail: bart@physics.iberiapac.ge ; gogber@physics.iberiapac.ge ).}
} \\
\vskip 2.5cm

{\Large \bf Abstract}\\
\vskip 0.5cm

\quotation
{\small
   We investigate the gravitational behavior of spherical domain walls
   (bubbles) arising during the phase transitions in the early Universe.
   In the thin-wall approximation we show  the existence of the new solution
   of Einstein equations with negative gravitational mass of bubbles and the
   reversed direction of time  flow  on the shell. This walls exhibit
   gravitational repulsion just as the planar walls are assumed to do. The
   equilibrium radius and critical mass of such objects are found for
   realistic models.  }
\endquotation
\end{center}

\vskip 2.5cm
\section {Introduction.}

   \rm ~~~~Topological structures such as domain walls, strings and monopoles
could be produced at  phase transitions in the  Universe  as it cooled
\cite{BGV91,B88,CF75,L77,V85,V91}. Within the context of general relativity
they are assumed to be an unusual sources of gravity. Cosmic strings do not
produce any gravitational force on the surrounding matter locally, while
global monopoles, global strings and planar domain walls are repulsive
\cite{BGV91,BKT87,HL90,HS88,IS84,L88,V87,V85,V91}.

    We shall consider domain walls, produced through the breakdown
of discrete symmetry. Their stress-energy is  composed  of  surface
density $ \sigma $ and strong tension $p$ in two spatial directions
with the magnitude \cite{V85,V91}

\begin{equation}
\sigma  = - p = const.
\label{0.1}
\end{equation}
This state equation corresponds to de Sitter expansion in the wall-plane and
the borders of the wall run away with the horizon. We can speak about
the gravitational field of the wall only in normal direction to the wall.
If one assumes that for such objects it is possible to use Newtonian
approximation with the mass described by Tolman's formula \cite{T69}

\begin{equation}
 M = \int (T_0^0 - T_1^1 - T_2^2 - T_3^3) \cdot \sqrt{-g}dV =
 \int (\sigma+2p) \cdot \sqrt{-g}dV = - \int \sigma \cdot \sqrt {-g}dV
\label{0.2}
\end{equation}
as it is usually assumed \cite{V85,V91}, then that tension acts as a
repulsive source of gravity and the planar domain wall has a negative
gravitational mass and exhibits repulsive gravitational field
\cite{BKT87,IS84,L88,V85,V91}.

    It is natural to think that the same  behavior  (gravitational repulsion)
must occur for the  spherical  domain  walls  (bubbles), since it is assumed
that they are described by the same state equations (1) (e.g. see
\cite{BKT87,IS84,L88}), different aspects of bubble-dynamics was
investigated also in papers
\cite{ABS91,ADLS84,ADLS85,AKMS87,APS89,BKT83,BKT832,BKT87,BKT87p,BGG87,C83,
C77,GHG90,HKLLM92,LM86,L77,L84,L88,MSSK82,SKSM82,T91}).
On the other hand, according to Birkhoff's theorem, the empty space,
surrounding any spherical body (including bubbles),  is  described  by
Schwarzschild metric. This metric contains parameter $m$ (corresponding to
the mass of gravitating body) which  is  described  through the integral over
energy density of the body

\begin{equation}
 m = \int T_0^0 \cdot \sqrt{-g}dV + const .
\label{0.3}
\end{equation}
While for planar domain walls (which stretch the horizon) the negative
gravitational mass (\ref{0.2}) can be admittable, for bubbles the
negativeness of mass (\ref{0.3}) from the first glance looks surprising since
$T_0^0$ is positively defined everywhere. Thus independently on the state
equation (\ref{0.1}) the mass (\ref{0.3}) usually is considered to be
positive \cite{LL75,MTW73,T69}.

However there can be no contradiction since in the case of spherical domain
walls (in difference to isolated matter for which the condition of
energodominance is valid) it is impossible to surround the full source by
any boundary inside the horizon (just as it is for planar domain walls). The
domain wall is only the "part" of scalar field solution which fills the
whole Universe up to horizon and which has nonzero vacuum expectation value
(VEV) even in infinity. The result is that the quantity
$\int T_{\mu\nu}\cdot dS^{\nu} $ is not a 4-vector of energy-momentum and one
can not define the energy simply as $\int T_{00}\cdot dxdydz $ \cite{SI48}.
For example, the energy density of expanding spherical domain wall remains
constant (see (\ref{0.1})) despite increasing of its surface i.e.  this
object "takes" the energy from vacuum, while the energy of vacuum depends on
VEV of scalar field in the whole space including infinity. This means that
for the case of topological objects we can not neglect the boundary terms at
infinity since the scalar field forming the wall does not vanish there. The
same situation is for global monopoles which also can exhibit the
gravitational repulsion \cite{HL90}.

The other solution of the negative mass problem can be the fact that the
domain walls are not described by the state equation (\ref{0.1}). One must
take into account the flux out from the volume of integration or some external
forces stabilizing the domain wall. As a result the state equation can have
a principally different form and both the spherical and planar walls can be
gravitationally attractive.

The other possible reason of disagreement may be that planar domain walls can
be described by the state equation (\ref{0.1}) while the bubbles can not.

In this paper we argue that if nevertheless one assumes that spherical
domain walls are described by the state equation (\ref{0.1}) (as it is
usually considered, see e.g.  \cite{BKT87,IS84,L88}), then
they must have a negative gravitational mass and must be repulsive
\cite{BG94}.

To start with, one has to note that Schwarzschild parameter $m$ in formula
(\ref{0.3})  contains  arbitrary integration constant which must be fixed
from the boundary conditions \cite{LL75,MTW73,T69}.  For the ordinary matter
the boundary  conditions both  in the center of the body and at the infinity
lead to  zero  value  of this constant. For the topological objects surface
singularities in gravitational quantities allow us to use only one boundary
condition dependently on the region we are interested. When investigating
the outer gravitational field  one has to use the boundary condition at a
large distance, where the gravitational potential must be expressed by
Newton's formula. Fixing the constant from this condition it is easy to see
that the mass of topological objects is determined by Tolman's formula
(\ref{0.2}) (on the other hand when investigating field inside the object
one must take $m = 0$, because otherwise the metric will have a singularity
at the center $r = 0$).  Thus the constant in formula (\ref{0.3}) takes into
account the pressure (which can not be negligible for topological objects)
and so the results from formulae (\ref{0.2}) and (\ref{0.3}) coincide.  For
the bubbles this formulae give negative value of $m$ and these objects
appear to be repulsive (as the planar walls do).  This leads us to
conclusion that the space around such physically admitable objects as
spherical domain walls are can be described by Schwarzschild solution with
negative parameter $m$ \cite{BG94}.

Here we would like to emphasize that though Tolman's formula is valid only
for static objects, nevertheless even for the expanding (or collapsing)
spherical domain walls the expression for active gravitational mass of the
 bubble (obtained, for example, in the thin-wall formalism, see sections 7
 and 9) must contain the term corresponding to the gravitational energy
(\ref{0.2}) and in static Newton's limit (when the velocity of bubble
expansion tends to zero) this expression must coincide with Tolman's formula.

  Since the exact solution of Einstein's equations for thick bubbles is
unknown, bubble dynamics usually is analyzed in the thin-wall formalism
\cite{C70,CI67,I66,K68,MTW73}. The above-mentioned problem emerged there too.
In this formalism it also was obtained that active  gravitational mass of
the spherical domain wall is positive, i.e. its gravitational field is
attractive \cite{BKT87,IS84,L88}. The disagreements in gravitational
properties of planar and spherical domain walls were explained by
instability of the latter \cite{IS84}, or by existence of positive energy
source stabilizing the bubble \cite{L88}.

We shall investigate the bubble dynamics in the thin-wall formalism and show
 that the space outside the bubble can be described by Schwarzschild
 solution with negative mass-parameter i.e.  spherical domain walls are
 repulsive. We shall see that this solution requires the reversal of time
 flow on the wall-surface. Note that the solutions of Einstein equations
where in different space-regions time flows in opposite directions are well
known. For example in Reissner-Nordstrom metric in the  region between the
upper and Cauchy radii the time coordinate changes its direction to the
opposite.

 Gravitational repulsion of bubbles can solve different paradoxes (for
 example the blueshift instead of redshift) appearing in models with large
 pressure \cite{G85}.

It is worth to mention another contradicting example  --- a planar domain
wall stretched by a static cosmic  string  hoop  \cite{IS84}.  Such system
must repel a test particle placed next  to  the  domain wall (the domain
wall is repulsive, while cosmic  string  does  not act gravitationally
\cite{V85,V91}), whereas for a distant observer it must behave as a bubble,
i.e. according to \cite{BKT87,IS84,L88} it must be gravitationally
attractive. This paradox can be solved only if bubbles are repulsive.

The time flow reversal can explain also the problems mentioned in
\cite{BGG87}. Consider the finite region of false vacuum with nonzero energy
density (and thus with negative pressure) separated by a domain wall from
an  infinite  region  of true vacuum with zero energy density. In this case
an observer placed into the false-vacuum region (described by de Sitter
metric)  would expect to see inflation and thus increasing of  bubble
radius.  At the same time an outer observer on the true-vacuum side
(described by Schwarzschild metric) would discover that  the  pressure
forces are inward and bubble must collapse, thus he would not see an
increase of the radius of curvature. This  problem  was  explained  by
assumption, that false-vacuum region does not move out  into  true-vacuum
region and this two areas expand separately \cite{BGG87}.  However,
if we take into account the time reversal in the region with the strong
pressure and the negative mass (i.e. the false vacuum inside the bubble) this
paradox can be explained in the frames of the standard scenario of phase
transitions, when the false-vacuum region expand into true vacuum region.

    In the next section we describe some features of domain walls.

    The section 3 is dedicared to review of the thin-wall formalism. The
motion equations of thin shells are given.

     In section 4 we consider the surface stress-energy tensors for thin
shells.

     In section 5 we write the motion equations for  spherical shells.

     In section 6 the sign ambiguity of motion equations is discussed. It is
shown the possibility of existence of a new solution of Einstein equations
for bubbles (Schwarzschild solution with negative mass) in a thin-wall
formalism.  The negative value of the bubble  mass  in the motion equations
leads to the time flow reversal on the bubble.

    In sec. 7  the simplest examples of spherical dust and domain walls in
vacuum are considered. It is shown that gravitational fields around this
objects are described by Schwarzschild metric with the opposite signs of
mass parameter.

    In sec. 8 some unusual properties  of  repulsive  spheres  are
mentioned. The embedding of Schwarzschild metric with different signs of
mass parameter in 6-dimensional space-time is investigated.

    In sec. 9 the dynamic of repulsive bubbles in general case of charged
    bubbles and nonzero vacuum energy density is
discussed. The equilibrium radius and critical mass  of  static bubbles
are found for different symmetry violation scales.

    In sec. 10 the problem of stability is considered.


\section{ The domain wall. }

{}~~~~~A first-order phase transitions which take place in the most of
cosmological models proceed through the nucleation of the new phase bubbles
(see e.g. \cite{B88,HKLLM92,L77}). Such processes take place at the very
beginning of the phase transition. At the final stage of transition the old
phase fragments which are left up to that moment also take the spherical
form (for example due to surface tension effects or dissipation). The
surface of phase separation in the case of the first-order phase transition
is represented by the so-called domain wall. Such objects are created when
the vacuum manifold for the order parameter or scalar field $\varphi$
driving the symmetry breaking has a discrete symmetry  \cite{V85,V91}. At
the time of transition there can be both infinite and closed surface walls.

To model the essential features of a domain wall, we will  consider a
scalar field with the Lagrangian

$$
L = \frac{1}{2} g^{\mu\nu}\partial _\mu\varphi\partial _\nu\varphi -
V(\varphi ) ,
$$
where $ V(\varphi )$ has two minima at nonzero $\varphi$ . Here the greek
indices take the values 0,1,2,3 and the metric has the signature
$(+,-,-,-)$. The example of such potential with discrete symmetry
$\varphi \rightarrow -\varphi$ is

$$
V(\varphi ) = -\frac{M_G^2}{2}\varphi^2 + \frac{\lambda}{4}\varphi^4 ,
$$
$M_G^2/\lambda$   representing the symmetry breaking scale and
$ \lambda $ --- the  self coupling constant.

  In the case of a flat wall in Minkowski space the  equations  of
motion

$$
\partial _\mu\partial_ \nu\varphi - \frac{\partial V(\varphi )}{\partial
\varphi} = 0
$$
admit the classical "domain wall" or "membrane"  solution,  depending on one
coordinate:

$$
\varphi (x) = \frac{M_G}{\sqrt{\lambda}}\cdot \tanh\left[
\frac{M_G}{\sqrt{2}} ( x - x_0 ) \right] ,
$$
where $x$ is the normal direction to the wall and $x_0$ is the wall position.

In this case the VEV of scalar field
$<\varphi > \rightarrow \pm <\varphi_0  > = \pm M_G/\sqrt{\lambda}$ on
either side of the  wall and thus there is defined the wall surface $\Sigma$
--- the  three  dimensional surface on which $\varphi = 0$.

The energy-momentum tensor

$$
T_{\mu\nu} = \partial_\mu\varphi\cdot\partial_\nu\varphi - g_{\mu\nu}L
$$
for this solution takes the form :

\begin{eqnarray}
T_{00} = \left( \frac{\partial\varphi}{\partial x} \right)^2 =
\frac{M_G^4}{2\lambda}\cdot \cosh^{-4}\left[ \frac{M_G}{2} ( x - x_0 )
\right] ;
\nonumber \\
T_{xx} = 0 ; \nonumber \\
T_{yy} = T_{zz} = - T_{00}.
\label{2.6}
\end{eqnarray}

{}From the first equation of (\ref{2.6}) it is easy to see, that the
thickness and energy per unit area of this wall are respectively:

$$
\Delta \sim \frac{1}{M_G} ;
$$
$$
\sigma = \int\limits_{-\infty}^{+\infty} dx T_{00} \sim
\frac{M_G^3}{\lambda} .
$$

{}From the third equation of (\ref{2.6}) one can see that the wall solution
has a high pressure along the wall, which plays an essential role in
gravitational behavior of such objects (see Tolmans formula (\ref{0.2})).


\section{ Thin wall approximation. }

{}~~~~~In the limit of vanishing thickness of  the  domain  wall it can be
considered as an infinitely thin three-dimensional hypersurface $\Sigma$ with
the energy-momentum tensor $T^\mu_\nu$ having, in general, singularities on
it. This hypersurface divides the Riemannian manifold $^{(4)}V$ into two
parts $V^+$ and $V^-$. Each of them contains $\Sigma$  as a part of its
boundary.

    We intend to apply the Einstein equations

\begin{equation}
G_{\mu\nu} = R_{\mu\nu} - \frac{1}{2}\cdot g_{\mu\nu}R = 8\pi G\cdot
T_{\mu\nu}
\label{3.1}
\end{equation}
to such thin boundaries of phase separation and investigate their dynamics and
gravitational behavior using  the  thin-wall approximation. In (\ref{3.1})
$R_{\mu\nu}$ is the Ricci tensor, expressed in terms of Christoffel
connections $\Gamma^\lambda_{\mu\nu}$ in the usual way:

$$
R_{\mu\nu} = \partial_{[\rho}\Gamma_{\mu ]\nu}^\rho +
\Gamma^\sigma_{\rho [\sigma}\Gamma_{\mu ]\nu}^\rho
$$
(here the square brackets denote the antisymmetrization)
and
$$
\Gamma_{\mu\nu}^\lambda = \frac{1}{2}\cdot g^{\lambda\sigma}
( \partial_\nu g_{\sigma\mu} + \partial_\mu g_{\sigma\nu} -
\partial_\sigma g_{\mu\nu} ) ,
$$
$G = M_{Pl}^{-2}$ is the gravitational constant,
$M_{Pl} = 1,2\cdot 10^{19} GeV$ is the Planck mass, $\partial_\mu$ is usual
derivative.  (One must realize, that when speaking  about  a  thin shell we
always keep in mind that the thickness of the shell must not be smaller then
$1/M_{Pl}$, provided gravity is described just by the classical Einstein
equations).

We shall restrict ourselves to the case of time-like hypersurfaces $\Sigma$
with the energy-momentum tensor having singularities no stronger, then those
given by $\delta$-functions. Then the first derivatives of $g_{\mu\nu}$ are
discontinuous on $\Sigma$, while metric itself is still continuous on
$\Sigma$.  Gravitational formalism for such boundaries is considered in
details in papers \cite{BKT87,C70,CI67, I66,K68,MTW73}.

Let $\{ y^\mu \}^+\quad  (\{ y^\mu \}^- )$ be an arbitrary coordinate
system  in  $V^+\; (V^-)$ region, and $\{\xi^i\}$ be an arbitrary
coordinate system on $\Sigma$ (here the latin indices run through 0,1,2).
The coordinate charts $\{ y^\mu\}^+$ and $\{ y^\mu\}^-$ need not join
smoothly on $\Sigma$ .

    Let us assume, that equation of hypersurface $\Sigma$ in the  chosen
coordinates $y^\mu$ has the form

\begin{equation}
F(y^\mu ) = 0
\label{4.1}
\end{equation}
(Since all equations in the regions $V^+$ and $V^-$ differ only by the
indices "$+$" and "$-$", we, for simplicity, will not write them separately
if not necessary).

We can introduce the new function
\begin{equation}
n(y^\mu ) = \varepsilon \frac{F(y^\mu
)}{\sqrt{g_{\lambda\nu}\partial^\lambda F\cdot\partial^\nu F}}
\label{4.2}
\end{equation}
which describes the surface $\Sigma$. Here the sign function $\varepsilon$
depends on the orientation of the (1+2)-surface $\Sigma$.
If a displacement vector $dy^\mu$ lies on the hypersurface $n
= const$, then

$$
dn = \partial_\mu n \cdot dy^\mu = 0  ,
$$
$\partial_\mu$ being the usual derivative. Therefore a vector

\begin{equation}
N_\mu \equiv \partial_\mu n |_\Sigma
\label{4.3}
\end{equation}
is the unit normal vector to the hypersurface $\Sigma$ :

$$
N_\mu N^\mu = - 1
$$
(recall that we consider the timelike $\Sigma$ ).

We can also introduce the unit vectors, tangential to $\Sigma$ :

$$
e^\mu_i = \frac{\partial y^\mu}{\partial \xi^i}.
$$
The tetrad field $(N^\mu, e^\mu_i)$ is thereby defined on $\Sigma$ .

For our further calculations it is convenient  to  split  the field
equations into their components orthogonal and tangential to the wall
surface $\Sigma$ \cite{BKT87,C70,CI67,I66,K68,MTW73}.

For this purpose the interval, which in $y^\mu$ chart has the form

$$
ds^2 = g_{\mu\nu}dy^\mu dy^\nu
$$
($g_{\mu\nu}$ is the metric tensor, which determines the geometry in $V$
region), can be written in Gaussian normal coordinates in the form

\begin{equation}
ds^2 = -dn^2 + \gamma_{ij}(\xi^k, n)d\xi^id\xi^j .
\label{4.7}
\end{equation}
{}From (\ref{4.1}) and (\ref{4.2}) it is clear, that $n = 0$ is the equation
of
hypersurface $\Sigma$ and thus the interval

$$
ds^2 = \gamma_{ij}(\xi^k)d\xi^id\xi^j
$$
determines the 3-geometry on $\Sigma$.

Any  vector  and  tensor  naturally  is  splitted   into   its components
orthogonal and tangential to $\Sigma$ :

$$
A^\nu = A^nN^\nu + A^ie^\nu_i,
$$
$$
Q^{\mu\nu} = Q^{nn}N^\mu N^\nu + Q^{ni}e^\mu_iN^\nu + Q^{jn}N^\mu e^\nu_j
+ Q^{ij}e^\mu_ie^\nu_j .
$$
Using the Gaussian coordinates the Einstein equations (\ref{3.1}) also can be
decomposed into scalar, 3-vector  and  3-tensor  parts  in  respect  to  the
coordinate transformations on Hypersurface $\Sigma$ :

\begin{eqnarray}
G^n_n = - \frac{1}{2}\cdot\left( ^{(3)}R + (K^i_i)^2 - K^i_lK^l_i \right) =
8\pi G\cdot T^n_n  ; \nonumber \\
G^n_i = D_mK^m_i - D_iK^l_l = 8\pi G\cdot T^n_i ; \nonumber \\
G^i_j = ^{(3)}G^i_j - N^\mu D_\mu (K^i_j - \delta^i_jK^l_l) +
K^l_lK^i_j - \nonumber \\
\frac{1}{2}\cdot \delta^i_j \left( (K^l_l)^2 + K^l_mK^m_l
\right) = 8\pi G\cdot T^i_j .
\label{4.11}
\end{eqnarray}
Here the $D_\mu$ denotes covariant differentiation with  respect of the
connection in $V$, while $D_i$ denotes  3-dimensional covariant
differentiation with respect to  the  connection  on $\Sigma$ , $ ^{(3)}R$
is three dimensional scalar curvature, $ ^{(3)}G^i_j$ is three dimensional
Einstein tensor and $K^i_j$ is the extrinsic curvature tensor of the
hypersurface $\Sigma$ , which is defined in the following way:

\begin{equation}
K_{ij} = - e^\mu_ie^\nu_j D_\nu N_\mu  .
\label{4.12}
\end{equation}
In  the  equations  (\ref{4.11})  we  have  used   the   well-known
Gauss-Kodazzi equations \cite{BKT87,C70,CI67,I66,K68,MTW73}.

    Note, that in the Gaussian coordinates (\ref{4.7}) the  tensor  of
extrinsic curvature (\ref{4.12}) acquires the simple form:

\begin{equation}
K_{ij} = -\Gamma^n_{ij} = -\frac{1}{2}\cdot \partial_n\gamma_{ij}  .
\label{4.13}
\end{equation}

Since the 3-geometry on the surface $\Sigma$ is, by assumption, well
defined, the components of Christoffel symbols (in  the  intrinsic
coordinates $\xi^i$), not containing indices $n$ are regular.  Components
with two or three indices $n$ are equal to zero, while with one $n$ --- are
discontinuous and have the step-function behavior when crossing $\Sigma$.
However $g_{\mu\nu}$ is assumed to be regular on $\Sigma$,  and
$g^+_{\mu\nu}$ and $g^-_{\mu\nu}$ have to match continuously on the shell.
Three-curvature tensor $ ^{(3)}R_{ij}$  of hypersurface $\Sigma$ also does not
contain singularities and is expressed  in terms of 3-metric tensor
$\gamma_{ij}$ in the usual way just as $R_{\mu\nu}$ is  expressed  by
$g_{\mu\nu}$.

Integrating the $(ij)$ component of equation (\ref{4.11}) in the normal
direction by the proper distance $dn$ through $\Sigma$ we can get the
so-called Lanczos equation (see \cite{BKT87,C70,CI67,I66,K68,MTW73}).

\begin{equation}
\left[ K^i_j \right] - \delta^i_j \left[ K^l_l \right] =
8\pi G\cdot S^i_j ,
\label{4.15}
\end{equation}
where

$$
\left[ K^i_j \right] \equiv lim_{\epsilon \rightarrow 0 }
\left( K^i_j (n=+\epsilon ) - K^i_j (n=-\epsilon ) \right)
$$
is the discontinuity of the outer curvature tensor and
$$
S^i_j = lim_{\epsilon \rightarrow 0 } \int\limits_{-\epsilon}^{+\epsilon}
T^i_j dn
$$
is the intrinsic surface energy tensor  on  the  $\Sigma$.

Noting (\ref{4.13}), writing the $(nn)$ and $(ni)$ components of Einstein
equations in the regions $V^+$ and $V^-$ and subtracting the corresponding
equations for the $V^+$ region from those for the $V^-$ region one yields
(using (\ref{4.15})):

\begin{eqnarray}
 D_jS^j_i + \left[ T^n_i \right] = 0 \nonumber \\
\{ K^i_j \} S_i^j + \left[ T^n_n \right] = 0,
\label{4.18}
\end{eqnarray}
where
 $\{ K_j^i \} = \frac{1}{2} \cdot lim_{\epsilon \rightarrow 0 }
\left( K^i_j (n=+\epsilon ) + K^i_j (n=-\epsilon )\right) $.

Now to describe completely the gravitational behavior and dynamics of bubbles
one needs to investigate the outer curvature and energy-momentum tensors of
these objects, insert these quantities into motion equations (\ref{4.15})
and (\ref{4.18}) and solve them.

\section{ The surface stress-energy tensor.}

{}~~~~~We shall restrict ourselves to a pure vacuum case, i.e.  when
there are no particles at the either sides off the shell.

    An observer, which moves with an element of the  shell  finds that the
momentum of the matter all the time lies on  the  surface of the shell i.e.
$$
T_{\mu\nu}\cdot N^\mu = 0
$$
on $\Sigma$  and

$$
T_{\mu\nu} = 0
$$
outside $\Sigma$ . Thus in the intrinsic coordinates

\begin{equation}
T_n^n = T_n^i = 0
\label{5.3}
\end{equation}
and the surface energy-momentum tensor for an  observer  on  $\Sigma$   is
represented by the tensor $S^i_j$.

It is easy to see that with the condition (\ref{5.3}) the motion equations
(\ref{4.18}) are satisfied automatically (they become identities) and one is
left only with Lanczos equations (\ref{4.15}).

Now let us suppose, that the surface energy-momentum tensor has the
structure of an ideal fluid

\begin{equation}
S^{ij} = (\sigma + p )u^iu^j - p\gamma^{ij} ,
\label{5.4}
\end{equation}
where $u^i$ is the 3-velocity of an element of hypersurface $\Sigma$ ,  while
$\sigma$  and  $p$ are the surface energy density (energy per unit area) and
surface tension respectively.

    The energy-momentum conservation in the intrinsic coordinates
implies that equation

\begin{equation}
D_iS^{ij} = 0
\label{5.5}
\end{equation}
expresses the energy-momentum balance of matter on the shell.  For
tensor (\ref{5.4}) this conservation equation becomes

\begin{equation}
u^j\cdot D_i\left[ (\sigma + p )u^i \right] + (\sigma + p)u^iD_iu^j -
\gamma^{ij}\cdot D_ip = 0  .
\label{5.6}
\end{equation}
Multiplying (\ref{5.6}) by $u_j$ we yield

\begin{equation}
D_i\left[ (\sigma + p)u^i \right] - u^i\cdot D_ip = 0.
\label{5.7}
\end{equation}
Feeding the last relation back into (\ref{5.6}) we obtain

\begin{equation}
(\gamma^{ij} - u^iu^j)\cdot D_ip - (\sigma + p)u^iD_iu^j = 0 .
\label{5.8}
\end{equation}

    The conservation equations (\ref{5.7}) and (\ref{5.8}) are easily
solved in two cases.

    (a). For the dust wall

$$
p = 0
$$
and we have

$$
D_i(\sigma u^i) = 0 ,
$$
that states that the total amount of dust is conserved.

    (b). For the domain wall

$$
p = -\sigma
$$
and from (\ref{5.7}) it follows immediately that

$$
\sigma = const  .
$$

This two examples are considered in section 7 in spherically symmetrical
case.

\section{ The geometry of spherical vacuum shells.}

{}~~~~~Here we consider the bubbles of spherical form. For the spherical
shells the Lanczos equations (\ref{4.15}) (the only nontrivial equation
of motion for the vacuum case) takes the form (see \cite{BKT87}):

\begin{eqnarray}
 \left[ K^2_2 \right] = 4\pi GS^0_0 ,    \nonumber  \\
 \left[ K_0^0 \right] + \left[ K^2_2 \right] = 8\pi GS^2_2 .
\label{7.6}
\end{eqnarray}

For the spherical shell the most convenient  choice  for  the outer region
coordinates $y^\mu$ is the ordinary  spherical  coordinate system. The
symmetry features of the problem allows us to  choose the coordinates
$\vartheta$ and $\varphi$ to be continuous across $\Sigma$ , i.e.

$$
\vartheta^+ = \vartheta^- = \vartheta ;
$$
$$
\varphi^+ = \varphi^- = \varphi .
$$

    The metric off the shell in the $V^\pm$ regions must be the solution
of spherically symmetrical Einstein equations. According to Birkhoff's
theorem such metric has the form \cite{LL75,MTW73,T69}

\begin{equation}
(ds^\pm )^2 = f^\pm \cdot (dt^\pm )^2 - \frac{1}{f^\pm }\cdot (dr^\pm )^2 -
(r^\pm )^2\cdot d\Omega^2 ,
\label{6.2}
\end{equation}
where

$$
d\Omega^2 = d\vartheta^2 + \sin^2\vartheta\cdot d\varphi^2.
$$

    For the intrinsic coordinates $\xi^i$  of the shell  we  shall  use
the proper time $\tau$  and the spherical angles $\vartheta$ ,$\varphi$ .
Then the metric on (1+2)-surface $\Sigma$ , induced both by exterior
$(ds^+)^2$ and interior $(ds^-)^2$ metrics (\ref{6.2}), can be expressed as

\begin{equation}
ds^2 = d\tau^2 - R^2(\tau )d\Omega^2 ,
\label{6.6}
\end{equation}
where $R(\tau )$ is the shell radius.

    Generally speaking, the time and radial  coordinates  are  not
continuous on the shell, but some restrictions on these coordinates is
obtained from the junction of metrics (\ref{6.2}), (\ref{6.6}):

\begin{equation}
(ds^+)^2 = (ds^-)^2 = ds^2_{|\Sigma} .
\label{6.7}
\end{equation}
The radius of the shell $R(\tau )$ can be described  in  the  coordinate
invariant way, so the (\ref{6.7}) gives only two conditions.  Identification
of the two-spheres $(r,t = const)$ on $\Sigma$  yields

\begin{equation}
r^+ = r^- = R(\tau )_{|\Sigma } ,
\label{6.8}
\end{equation}
while the time  coordinate  can  be  discontinuous  on  the  shell:
$(t^+ \neq t^-)_{|\Sigma}$. Comparison of timelike lines
$(\varphi ,\vartheta  = const)$ gives

\begin{equation}
d\tau^2 = f^+\cdot (dt^+)^2 - \frac{1}{f^+}\cdot dR^2 = f^-\cdot (dt^-)^2 -
\frac{1}{f^-}\cdot dR^2 .
\label{6.9}
\end{equation}
{}From these relations it is easy to find that

$$
\left( f^\pm \dot t^\pm \right)^2 = f^\pm +  \dot R^2 ,
$$
\begin{equation}
\frac{1}{f^\pm }\cdot \left( \frac{dR}{dt^\pm } \right) ^2 = f^\pm -
\frac{1}{(\dot t^\pm )^2}
= \frac{f^\pm \cdot \dot R^2}{f^\pm + \dot R^2},
\label{6.11}
\end{equation}
where the overdot denotes the derivation with respect of  proper  time
$\tau$ . Thus, after the procedure of junction (\ref{6.7}), there remains
only one unknown function $R(\tau )$ which must obey the  Einstein
equations on the shell.  It means that from the two equations of motion
(\ref{7.6}) only one is independent. We shall choose and solve the first
one (then the second one will be satisfied automatically).  Now we have to
find the outer curvature tensor.

It is easy to see, that equation (\ref{6.8}) is the equation of motion of
the shell (compare with (\ref{4.1})):

$$
F = r - R(\tau ) = 0 .
$$
Noting, that

$$
dF/dt = - dR/dt;
$$
$$
\quad dF/dr = 1
$$
and using equations (\ref{6.2}),(\ref{6.11}) we can compute the outer
normal to $\Sigma$ (see (\ref{4.2}) and (\ref{4.3})):

$$
N_0 = - \left( \frac{dR}{dt} \right) \cdot N_1 ,
$$
$$
N_1 = \varepsilon \cdot |\dot t| ,
$$
$$
N_2 = N_3 = 0 .
$$
Here $\varepsilon$ is a sign function, which depends on the direction of an
outer normal $N_\mu$, i.e. on the orientation of (1+2)-surface $\Sigma$:

\begin{equation}
\varepsilon = sign\left( \dot t, \frac{\partial r}{\partial q} \right),
\label{e}
\end{equation}
where $q$ is any coordinate increasing along the direction from the
bubble-center.

  Using formulae (\ref{4.12}) one can find the components of  extrinsic
curvature tensor:

\begin{eqnarray}
 K_0^0 = K^2_2 + \frac{R}{\dot R} \cdot \dot K^2_2 , \nonumber \\
K^2_2 = - \varepsilon \frac{|f\dot t|}{R} .
\label{7.4}
\end{eqnarray}
Note that the first equation of (\ref{6.11}) admits both signs for $f\dot
t$. The sign of $\varepsilon$ is equivalent to tha sign of $f\dot t$, thus
instead of the second equation (\ref{7.4}) one can write
\begin{equation}
K^2_2 = - \frac{f\dot t}{R} .
\label{7.41}
\end{equation}

Using (\ref{6.11}) one can yield from (\ref{7.4}):

$$
(K_\pm )^0_0 = -\varepsilon_\pm \frac{\ddot R + \frac{1}{2}\cdot (f^\pm)'
}{\sqrt{f^\pm + \dot R^2}} ,
$$
\begin{equation}
(K_\pm )^2_2 = - \frac{\varepsilon_\pm }{R}\cdot \sqrt{f^\pm + \dot R^2} ,
\label{7.5}
\end{equation}
where $f'$ denotes the derivation by $R$.

Thus we have shown that the dynamics of  spherically  symmetrical shell is
completely described by the first equation (\ref{7.6})

\begin{equation}
K_+ - K_- = 4\pi G \sigma ,
\label{8.1}
\end{equation}
where $\sigma = S_0^0$ is the energy density of the shell and $K$ is the only
independent component $K_2^2$ of the extrinsic curvature tensor described by
(\ref{7.41}) or (\ref{7.5}).

The fact that the first equation of (\ref{6.11}) admits both signs for
$f\dot t$ leads to the new class of solutions of Einstein equations for
bubbles with outer $\varepsilon_+ = -1$. This was missed in earlier works
(for example \cite{BKT87}), where it was considered that for bubbles, lying
above the Schwarzschild horizon, in outer space always $\varepsilon_+ = 1$.
The correct sign must be chosen from boundary conditions. We shall discuss
this subject below.

\section{ The sign ambiguity of extrinsic curvature. }

{}~~~~~Investigating the dynamics of vacuum spherical shells by means of
the motion equation (\ref{8.1}) one has to be careful when choosing the
sign of $\varepsilon$ and thus of $f\dot t$. For the given the inner and the
outer metrics $(ds^\pm)^2$ the $\varepsilon_\pm$ determines the global
geometry, i.e. how the inner geometry is stuck with the outer one. In
some cases the junction is impossible.

 As we have mentioned above, the sign of $\varepsilon$ depends on the
direction of an outer normal $N_\mu$, i.e. on the orientation of
(1+2)-surface $\Sigma$:

\begin{equation}
\varepsilon = sign\left( \dot t, \frac{\partial r}{\partial q} \right),
\label{e}
\end{equation}
where $q$ is any coordinate increasing along the direction from the
bubble-center.

Generally speaking, the sign of $\varepsilon$ may change at the shell motion,
for example, when the shell passes the horizon. If the initial conditions
allow the shell to collapse, then in the case $\varepsilon_+ > 0$ the shell
forms the black hole, while in the case $\varepsilon_+ < 0 $ --- the
warmhole.

In paper \cite{BKT87} the classification of possible geometries for different
signs of $\varepsilon_+$ and $\varepsilon_-$ was done. However in this paper
the dependence of the sign-function (\ref{e}) on the mutual orientation of
time coordinates on and off the shell, i.e. on the sign of $\dot t$ was
overlooked.  The bubble is the (1+2)-surface embedded in the 4-dimensional
space-time $^{(4)}V$ and not only the 2-sphere as it appears if we do not
consider the direction of time coordinate. Thus we get the more general
classification of bubble geometries compared to those presented in previous
papers.

  Generally speaking, we have four different cases of equation
(\ref{8.1}), depending on the signature of $\varepsilon_\pm$ or $(f \dot
t)^\pm$ and, therefore, on the signature of $K_\pm$.
The sign of $\varepsilon$ must be chosen from the motion equation
(\ref{8.1}) and the boundary conditions mentioned in the Introduction. One
has to take into account the positiveness of the surface energy density
$\sigma$ and the fact that there must not be the contradiction between the
Newton's limit of Einstein equations (\ref{8.1}) and Tolman's formula
(\ref{0.2}).

Let us discuss this problem and examine the signatures of $\varepsilon$
or $f\dot t$.

  a) The signature of $f^\pm$ ($\dot t_\pm > 0$).

Let us assume for a while that $\dot t_\pm > 0$. The the signature of $f\dot
t$ depends on the sign of $f$.
The general form of $f$ for the spherically symmetrical source is

\begin{equation}
f = 1 - \frac{2Gm}{r} + \frac{Ge^2}{r^2} - G\Lambda r^2 ,
\label{8.5}
\end{equation}
where $m$ is the mass of the source, $e$ is its charge,
$\Lambda = (8\pi /3) \cdot \rho $ and $\rho $  is
the vacuum energy density. In the simplest case

$$
e = \Lambda = 0
$$
and
\begin{eqnarray}
f_+ = 1 - \frac{2Gm}{R} ; \nonumber \\
f_- = 1 .
\label{8.7}
\end{eqnarray}
The signature of $f_+$ is positive if the radius of the shell lies above
the Schwarzschild horizon, i.e. when $R > 2m$ and is  negative  beyond
it. So, $f\dot t$ changes its signature, when the radius of  shell  passes
the horizon. The standart metric (\ref{8.5}) has an unphysical singularity
at the horizon caused by a poor  choice  of  the  coordinate system. To be
more correct in correspondence between the signature of $K$ and the radius
of shell, one can use another coordinate  system, for example the isotropic
coordinates (see \cite{BGG87}), which  have no singularities at the horizon.
In ref. \cite{BGG87} it was investigated (in Kruscal-Szekares coordinates)
the case when the value $f\dot t$ becomes negative for the bubbles crossing
the horizon, the systematical catalog of possible
solutions of the equation of motion was obtained and was shown, that two
signatures of $f$ correspond to the  two different, but equally acceptable
trajectories of shell.

  b). The signature of $\dot t_\pm$ ($f^{\pm} > 0$).

Here we shall examine the mostly interesting case of macroscopic bubbles
we shall assume, that the shell  always  lies  outside  the horizon  and
thus the unusual features of  the  space geometry are avoided. In this case
$f^\pm > 0$ and the signature of $f\dot t$ is determined by the sign of
$\dot t$.

Note that the example, when time flows in  opposite directions is the case
of Reissner-Nordstrom metric, when in the  region between the upper and
Cauchy radii the time coordinate changes its direction to the opposite. In
this case (in difference  to  the ours) the metric covers the whole
space-time $(r > 0)$ and such  feature was easy to notice.  The problem of
sign ambiguity entered due to taking a square root in equation (\ref{6.11}).
Similar situation is in Dirac theory when one has states with negative
energy that correspond to antiparticles.

We shall choose the positive direction for the 2-spheres in the way that the
radii increase in the direction of the outer normal. Then orientation of
$\Sigma$ is determined by the direction of time flow on the shell and

\begin{equation}
\varepsilon = sign \dot t
\label{eps}
\end{equation}

  It is clear, that the signature of $\dot t_\pm$ depends on the  direction
of time flowing in $V^\pm$ regions and on the shell. It is  natural  to
assume, that in $V^\pm$ regions time flows in the same direction i.e.

\begin{equation}
sign (t_+) = sign (t_-)
\label{8.8}
\end{equation}
while the direction of intrinsic time $\tau$  has to be found from the
equation of motion. From (\ref{eps}) and (\ref{8.8}) it is clear that
$\varepsilon_+ = \varepsilon_- \equiv \varepsilon = 1$ when $t_\pm$ and
$\tau$ flow in the same direction and $\varepsilon = -1$ in the opposite
case. Assumption (\ref{8.8}) leaves only two cases of equation (\ref{8.1}):
\begin{equation}
\sqrt{f_+ + \dot R^2} - \sqrt{f_- + \dot R^2} = \pm 4\pi\sigma GR
\label{em}
\end{equation}
which depend on the values of $f_\pm$.

  If $f_- > f_+$, than $| K_+ | < | K_- |$ and we have to take the lower
sign $(-)$ in (\ref{em}), because the energy density of the shell  is
positive (see (\ref{8.1})). According to (\ref{7.41}) and (\ref{8.8}) this
means  that $\dot t_\pm > 0$  and thus the time off and on the shell flows
in the same direction.

  Similarly if $f_- < f_+$ we have $\dot t_\pm < 0$, that means that the time
on the shall flows in the opposite direction to the  time  in  $V^\pm$
regions.

  Note, that for the simplest case (\ref{8.7}) the signature of $\dot t$
is connected with the sign of constant $m$, which is the integration
constant of solution of Einstein equations in $V^+$ region. This  constant
is fixed by matching of $f_+$ to the solution of Einstein  equations  on the
shell in the same coordinates. In the Newton approximation

$$
f_+ = 1 - 2G\cdot \int \left( T_0^0 - \frac{1}{2}\cdot T \right) \cdot
\frac{1}{R}\cdot dV  .
$$
The value and the signature of $m$ is determined by this formula and it can
be positive as well as negative dependently on boundary conditions.
In the next section we shall see that the correct choice of sign in
(\ref{em}) cancels the contradictions between Tolman's formula and Newton's
limit of (\ref{em}).

\section{Different kinds of shells}

{}~~~~~Here we shall consider two cases of macroscopic bubbles and show how
one has to choose the different signs of $\dot t$.

\medskip
{\large \bf a. The dust walls in vacuum.  }
\medskip

As we have mentioned above, for the shell of dust

\begin{equation}
p=0
\label{9.1}
\end{equation}
and the surface energy tensor of layer (\ref{5.4}) is equal to

$$
S^{ij} = \sigma u^iu^j  .
$$
Conservation law (\ref{5.7}) now has the form

\begin{equation}
D_i(\sigma u^i) = 0
\label{9.3}
\end{equation}
stating, that the total amount of dust is conserved. For the spherical shells
in the intrinsic coordinates $\xi^i$ with  the  interval
(\ref{6.6}) we have

$$
u^0 = 1 ;
$$
$$
u^1 = u^2 = 0 .
$$
Therefore

$$
D_iu^i = \Gamma^i_{0i} = \frac{2\dot R}{R} .
$$
So that equation (\ref{9.3}) reduces simply to

$$
\dot \sigma + 2\sigma \dot R/R = 0  .
$$
This equation is easily solved and yields

\begin{equation}
\sigma = \frac{const}{R^2} .
\label{9.7}
\end{equation}
To fix the constant in (\ref{9.7}) one can use  the  Tolman's  formula for
the proper mass of  the  shell  in  the  state  of  rest  (see (\ref{0.2})).
For the ideal fluid

\begin{equation}
M = 8\pi \cdot \int \left[ \sigma - \frac{1}{2}\cdot (\sigma - 2p) \right]
\cdot \delta (r-R)\cdot r^2dr = 4\pi R^2 \sigma = 4\pi \cdot const  .
\label{9.8}
\end{equation}
Using (\ref{9.1}) and (\ref{9.7}) for the $const$ from (\ref{9.8}) we obtain

$$
const = \frac{M}{4\pi} .
$$

In the simplest case of uncharged shells of dust in vacuum when the metric
is given by the formulae (\ref{8.7}) the motion equation (\ref{em}) takes
the form

\begin{equation}
\sqrt{1 - 2Gm/R + \dot R^2} - \sqrt{1 + \dot R^2} = \pm 4\pi \sigma G R =
\pm G\frac{M}{R} .
\label{9.10}
\end{equation}
We can rewrite (\ref{9.10}) in the form

\begin{equation}
m = \mp M\cdot \sqrt{1+\dot R^2} - G\frac{M^2}{2R} .
\label{9.12}
\end{equation}
The term $-GM^2/2R$ represents the gravitational interaction energy
according to the Newton's law, and $M\cdot \sqrt{1+\dot R^2}$ represents the
internal mass of the shell (the root is the analog of the Lorentz factor and
is equal to 1 in equilibrium, when $R = 0$ ).

{}From this relation it is easy to find that one has to choose in (\ref{9.10})
the {\it lower} sign "$-$" corresponding to $f\dot t > 0$ ($\varepsilon =
1$). Certainly, in this case $m > 0$, $f_+ < f_-$ and the condition of
surface energy density positiveness is satisfied. Besides, it is easy to see
that with this choice of sign the static Newton's limit of (\ref{9.12})
coinsides with Tolman's formula (\ref{9.8}) (though Tolman's formula is
valid only for static states, which, as we'll see, do not occur in this
simplest model, the term corresponding to gravitational energy of the shell
must enter the expression for the shell mass with the correct sign. Then in
static Newton's limit ($\dot R = 0; \quad G = 0$) the mass of the shell will
be described by Tolman's formula).

Note that if one would choose the {\it upper} sign "$+$" in (\ref{9.10})
then he should receive the contradiction between the static Newtons limit of
(\ref{9.12}) and Tolman's formula (\ref{9.8}).

Now let us show that in this simplest model there is no equilibrium state
for the dust shell. The conditions

\begin{equation}
\dot R = 0 , \quad \ddot R = 0
\label{9.13}
\end{equation}
must be satisfied for the spherical shell  to  be  in  equilibrium state.
The first of them leads from (\ref{9.12}) to the equation

$$
m = M - G\frac{M^2}{2R} .
$$
Taking $d/d\tau$ of (\ref{9.12}) we obtain

$$
\ddot R = \frac{1}{M^2}\cdot \left( m + G\frac{M^2}{2R} \right) \cdot
\frac{dm}{dR} ,
$$
so the second condition (\ref{9.13}) gives the stipulation

\begin{equation}
\frac{dm}{dR} = 0 .
\label{9.16}
\end{equation}
Using (\ref{9.12}) and (\ref{9.16}) one can see, that the radius of stability
of the shell $R_{stab} \rightarrow \infty$, i.e. there is no equilibrium
configuration for the dust wall in vacuum.

\medskip
{\large \bf b. Domain wall in vacuum.}
\medskip

In the case of domain wall one has the picture opposite to the previous
case.  For the domain wall $$p =\sigma$$ and the surface energy tensor of
layer (\ref{5.4}) is

$$
S^{ij} = \sigma\gamma^{ij} .
$$
{}From the conservation law (\ref{5.7})

$$
D_i\sigma = 0
$$
we obtain

$$
\sigma = const .
$$
{}From the Tolman's formula (\ref{0.2}) we have:

\begin{equation}
M = 8\pi \cdot \int \left[ \sigma - \frac{1}{2}\cdot (\sigma - 2p) \right]
\cdot \delta (r-R)\cdot r^2dr = -4\pi\sigma R^2  .
\label{10.4}
\end{equation}
One can see that the proper mass of layer is negative, if the energy density
of domain wall $\sigma$  is positive.

Considering again the simplest case (\ref{8.7}) one obtains the motion
equation
\begin{equation}
\sqrt{1- \frac{2Gm}{R} +\dot R^2} - \sqrt{1 + \dot R^2} = \pm 4\pi \sigma GR
= \mp G\frac{M}{R},
\label{10.6}
\end{equation}
where $M$ is described by (\ref{10.4}).

Again let us rewrite (\ref{10.6}) in the form
\begin{equation}
m = \pm M\cdot \sqrt{1+\dot R^2} - G\frac{M^2}{2R} .
\label{10.7}
\end{equation}
This formula as well as (\ref{9.12}) corresponds to the energy balance
equation and must contain the Newton's gravitational energy of the shell.

In opposite to dust-wall case, now one has to choose in (\ref{10.6}) the
{\it upper} sign "$-$" corresponding to $f\dot t < 0$ ($\varepsilon = -1$).
Certainly, in this case $m < 0$, $f_+ > 1$ and the surface energy density
positiveness condition is satisfied. Besides, for this choice of sign the
static Newton's limit of (\ref{10.7}) coinsides with Tolman's formula
(\ref{10.4}).

We emphasize once more that though Tolman's formula is valid only for static
states, the term corresponding to gravitational energy of bubble must enter
the expression for the bubble mass with the correct sign. Then in static
Newton' limit the mass of the bubble will be described by Tolman's formula.

One can see, that when one chooses the upper sign in (\ref{10.6}) then

$$
m = - \left( 4\pi\sigma R^2 \sqrt{1+\dot R^2} +
8\pi^2\sigma^2GR^3 \right).
$$
In this expression the mass is negative and we can  believe,  that distant
observer will be repelled by the domain wall.

  If one considers the equation (\ref{10.6}) with different sign in the
right side (see \cite{BKT87,IS84,L88}), one will obtain the positive value
for mass

$$
m =  \kappa R^2 \sqrt{1+\dot R^2} - \frac{G\kappa^2R^3}{2}.
$$
This expression contains the rest mass (\ref{10.4}) with the  wrong  sign
(the first term with $\dot R  = 0$) and contradicts with Tolman's formula
(\ref{10.4}). Thus Tolman's formula helps us to choose the correct sign in a
motion equation (\ref{10.6}).

   Thus the correct choice of the sign in the motion equation (\ref{10.6})
leads to the new solution of Einstein equations for bubbles (Schwarzschild
solution with negative mass) which shows that spherical  domain  walls are
repulsive. This fact requires a further investigation of bubble-dynamics and
their creation in cosmological models.

\section{ Schwarzschild space with negative mass.}

{}~~~~~Let us investigate the properties of Schwarzschild metric

\begin{equation}
ds^2 = \left( 1 - \frac{b}{r} \right) \cdot dt^2 - \frac{1}{1-b/r}
\cdot dr^2 - r^2 \cdot (d\theta^2 + \sin^2\theta \cdot d\varphi ^2 )
\label{11.1}
\end{equation}
in the case, when constant $b$ (which is related to the active  gravitational
mass of the source), can be negative.

First of all, for such spaces $g_{00} > 1$ and the velocity of light exceeds
its velocity measured in Minkowski space. However this fact does not cause
appearance of takhions.

For the metric (\ref{11.1}) the nonzero components of curvature tensor are

$$
R_{trtr} = \frac{b}{r^3} ,
$$
$$
R_{t\theta t\theta} = \frac{R_{t\varphi t\varphi}}{\sin^2\theta} =
-\frac{b(r-b)}{2r^2} ,
$$
$$
R_{r\theta r \theta} = \frac{R_{r\varphi r\varphi}}{\sin^2\theta} =
-\frac{b}{2(r-b)} ,
$$
$$
R_{\theta\varphi\theta\varphi} = - br\cdot \sin^2\theta .
$$
In this case the complex invariants of gravitational field have the form

$$
I_1 = \frac{1}{48} \cdot R_{\alpha\beta\gamma\delta}\cdot
\left( R^{\alpha\beta\gamma\delta} - \frac{i}{2}\cdot
\varepsilon^{\alpha\beta\mu\nu} \cdot R^{\gamma\delta}_{\mu\nu} \right) =
\left( \frac{b}{2r^3} \right) ^2 ,
$$
$$
I_2 = \frac{1}{96} \cdot R_{\alpha\beta\mu\nu}R^{\mu\nu\rho\sigma} \cdot
\left( R^{\alpha\beta}_{\sigma\rho} - \frac{i}{2}\cdot
\varepsilon_{\sigma\rho\lambda\kappa}\cdot R^{\lambda\kappa\alpha\beta}
\right) = -\left( \frac{b}{2r^3} \right) ^3 .
$$
One can see, that for the negative $b$ the second invariant changes its
sign. The result of this fact is that the  3-space  $(t = const )$ Gauss
curvature for the plains, normal to radius also changes  its sign:

\begin{equation}
k = \frac{P_{\theta\varphi\theta\varphi}}{\gamma_{\theta\theta}\cdot
\gamma_{\varphi\varphi}} = - P^r_r = \frac{b}{r^3} .
\label{11.4}
\end{equation}
Here $P_{\theta\varphi\theta\varphi}$, $P^r_r$, $\gamma_{\theta\theta}$,
$\gamma_{\varphi\varphi}$ are 3-curvatures and 3-metric  tensors of space
(\ref{11.1}) in the case $t = const$. Equation (\ref{11.4})  means, that for
the different signs of $b$ this subspaces belong to  the different types of
surfaces. To show this let us  embed  metric (\ref{11.1}) into Euclidean
space with more then 4 dimensions.

  An embedding of the linear element (\ref{11.1})  in  5  dimensions  is
impossible (see \cite{F59,R65}). Embedding of (\ref{11.1}) with positive
$b$ in  6 dimensions with signature 2+4

$$
ds^2 = dz_1^2 + dz_2^2 - dz_3^2 - dz_4^2 - dz_5^2 - dz_6^2
$$
is given by

\begin{eqnarray}
z_1 = \left( 1 - \frac{b}{r} \right) \cdot \cos t , \nonumber \\
z_2 = \left( 1 - \frac{b}{r} \right) \cdot \sin t , \nonumber \\
z_3 = \int \left( \frac{b(b+4r^3)}{4r^3(r-b)} \right) ^{1/2} \cdot dr ,
\nonumber \\
z_4 = r \cdot \cos \theta , \nonumber \\
z_5 = r \cdot \sin \theta \cdot \cos \varphi, \nonumber \\
z_6 = r \cdot \sin \theta \cdot \sin \varphi .
\label{11.6}
\end{eqnarray}
It is possible to eliminate the coordinates

$$
z_1^2 + z_2^2 = 1 - \frac{b}{r} ,
$$
$$
z_3 = \int \left( \frac{b(b+4r^3)}{4r^3(r-b)} \right) ^{1/2} \cdot dr ,
$$
$$
z_4^2 + z_5^2 + z_6^2 = r^2 .
$$
We note, that this surface in $z_1z_2$ plane is the 1-sphere,  in  the
$z_4z_5z_6$ is the 2-sphere and $z_3$ is space-like. Time-like  coordinates
$z_1$ and $z_2$ are periodic functions of $t$ so that embedding (\ref{11.6})
identifies distinct points of the original manifold. This suggests replacing
the trigonometrical functions by the hyperbolic functions and embedding of
(\ref{11.1}) for the positive $b$ is possible for the signature 1+5:

$$
ds^2 = dz_1^2 - dz_2^2 - dz_3^2 - dz_4^2 - dz_5^2 - dz_6^2 ,
$$
where
\begin{eqnarray}
z_1 = \left( 1 - \frac{b}{r} \right) \cdot \sinh t , \nonumber \\
z_2 = \left( 1 - \frac{b}{r} \right) \cdot \cosh t , \nonumber \\
z_3 = \int \left( \frac{b(4r^3-b)}{4r^3(r-b)} \right) ^{1/2} \cdot dr ,
\nonumber \\
z_4 = r \cdot \cos \theta , \nonumber \\
z_5 = r \cdot \sin \theta \cdot \cos \varphi, \nonumber \\
z_6 = r \cdot \sin \theta \cdot \sin \varphi .
\label{11.9}
\end{eqnarray}
We note that in $z_1z_2$ plane this surface now is hyperbola

$$
z_2^2 - z_1^2 = \left( 1 - \frac{b}{r} \right) .
$$

  For the negative mass
$$
   b  < 0
$$
we can see from (\ref{11.6}) and (\ref{11.9}), that coordinate $z_3$ became
complex and embedding (\ref{11.6}) now takes place in the space with
signature 3+3

$$
ds^2 = dz_1^2 + dz_2^2 + dz_3^2 - dz_4^2 - dz_5^2 - dz_6^2
$$
where

$$
z_1 = \left( 1 - \frac{b}{r} \right) \cdot \cos t ,
$$
$$
z_2 = \left( 1 - \frac{b}{r} \right) \cdot \sin t ,
$$
$$
z_3 = \int \left( \frac{b(b-4r^3)}{4r^3(r-b)} \right) ^{1/2} \cdot dr ,
$$
$$
z_4 = r \cdot \cos \theta ,
$$
$$
z_5 = r \cdot \sin \theta \cdot \cos \varphi,
$$
$$
z_6 = r \cdot \sin \theta \cdot \sin \varphi .
$$
Now the coordinate $z_3$ became time-like. Embedding (\ref{11.9}) is
possible for the signature 2+4

$$
ds^2 = - dz_1^2 + dz_2^2 + dz_3^2 - dz_4^2 - dz_5^2 - dz_6^2
$$
where

$$
z_1 = \left( 1 - \frac{b}{r} \right) \cdot \sinh t ,
$$
$$
z_2 = \left( 1 - \frac{b}{r} \right) \cdot \cosh t ,
$$
$$
z_3 = \int \left( \frac{b(b-4r^3)}{4r^3(r-b)} \right) ^{1/2} \cdot dr ,
$$
$$
z_4 = r \cdot \cos \theta ,
$$
$$
z_5 = r \cdot \sin \theta \cdot \cos \varphi,
$$
$$
z_6 = r \cdot \sin \theta \cdot \sin \varphi .
$$
In this case the coordinate $z_1$ is space-like and we can not  identify the
surface in $z_1z_2z_3$ space. $z_4z_5z_6$ surface is  the  2-sphere $r = R$
in all the cases and is identified in our work with the surface of domain
bubble.

To conclud this section we can say that the space (\ref{11.1}) with $b < 0$
can be embedded into a 6-dimensional space with signature (2+4) or (3+3),
while the usual Schwarzschild space $(b > 0)$ --- into the space with
signature (1+5) or (2+4) respectively.

\section{ The model.}

{}~~~~After we have formulated our idea \cite{BG94} how to avoid the
disagreements mentioned at the beginning of this paper let  us  investigate
a more general case of a spherically symmetrical charged bubble in vacuum,
when the metric outside the bubble is

$$
f_+ = 1 - \frac{2Gm}{r} + \frac{Ge^2}{r^2} - G\Lambda_+r^2 ,
$$
while inside ---

$$
f_- = 1 - G\Lambda_-r^2 ,
$$
where $\Lambda_\pm \equiv \frac{8\pi}{3}\cdot\rho$ , $\rho$
being the vacuum energy density in $V^\pm$ regions, and $e$ is the charge on
the shell.

  Now the motion equation (\ref{8.1}) takes the form
\begin{equation}
\sqrt{\dot R^2 + 1 - \Lambda_+GR^2 - \frac{2Gm}{R} + \frac{Ge^2}{R}} -
\sqrt{\dot R^2 + 1 - \Lambda_-GR^2} = \pm G\kappa R ,
\label{26}
\end{equation}
where

$$
\kappa \equiv 4\pi \sigma .
$$
Finding $m$ from this equation we yield:

\begin{equation}
m = \frac{\Lambda_- - \Lambda_+}{2}\cdot R^3 - \frac{G\kappa^2}{2}\cdot R^3
+ \frac{e^2}{2R} \mp \kappa R^2\cdot \sqrt{\dot R^2 + 1 - \Lambda_-GR^2}
\label{27}
\end{equation}
It is easy to understand the meaning of the terms in (\ref{27}). The first
term is the volume energy of the bubble (a difference between the old and
new vacuum energy densities). The second term  represents an energy of
gravitational self-interaction of  the  shell (the surface-surface binding
energy). The third term is the  electrostatic energy lying in the
three-space outside the bubble. The last term contains the kinetic energy of
the shell  and  surface-volume binding energy. As we have mentioned above,
we have to choose  the upper sign in equations (\ref{26}) and (\ref{27}) to
avoid the disagreements with Tolman's formula (\ref{10.4}).

  To examine dynamics of the bubble let us rewrite  the  equation
(\ref{27}) in the following way:

\begin{equation}
\dot R^2 - \Biggl[ \frac{1}{\kappa^2}\cdot\biggl( -\frac{m}{R^2} -
\frac{a}{2}\cdot R + \frac{e^2}{2R^3}\biggr)^2 + \Lambda_-GR^2 \Biggr] = - 1 ,
\label{28}
\end{equation}
where $a \equiv \Lambda_+ - \Lambda_- + G\kappa^2$. It is worth to note that
in this  equation the sign ambiguity disappears due to squaring. Introducing
new  dimensionless variables

\begin{eqnarray}
z \equiv \frac{R}{(-2m)^{1/3}} \cdot
\bigl( a^2 + 4\kappa^2\Lambda_-G \bigr)^{1/6} ,   \nonumber\\
\tau' \equiv \frac{\tau}{2\kappa}\cdot\bigl( a^2 + 4\kappa^2\Lambda_-G
\bigr)^{1/2}
\label{29}
\end{eqnarray}
and dimensionless parameters

$$
A \equiv a\cdot
\bigl( a^2 + 4\kappa^2\Lambda_-G \bigr)^{-1/2}  ,
$$
$$
E \equiv - \frac{4\kappa^2}{(-2m)^{2/3}} \cdot
\bigl( a^2 + 4\kappa^2\Lambda_-G \bigr)^{-2/3}  ,
$$
$$
Q^2 \equiv \frac{e^2}{(-2m)^{4/3}} \cdot
\bigl( a^2 + 4\kappa^2\Lambda_-G \bigr)^{1/6}  ,
$$
we can represent the motion equation (\ref{28}) as

$$
\biggl( \frac{dz}{d\tau'} \biggr)^2 + U(z) = E ,
$$
which is identical to that of the point like particle  with  energy $E$ ,
moving in one dimension under the influence of the potential

\begin{equation}
U(z) = - \Biggl[ z^2 - \frac{2A}{z}\cdot\biggl(1+\frac{Q^2}{z}\biggr) +
\frac{1}{z^4}\cdot\biggl(1+\frac{Q^2}{z}\biggr)^2\Biggr] ,
\label{32}
\end{equation}
For real trajectories $U$ must be negative since $E < 0$. Such potentials
(but for the case of uncharged shells $(Q = 0)$ with $ m > 0$) were
discussed in \cite{AKMS87,BGG87}.

  In the equilibrium state $\dot z_{|z=z_0} = 0$, where $z_0$ is the
equilibrium point,  $U(z_0) = E$ and one can find the critical mass of the
bubble:

\begin{equation}
m_0 = - \frac{4\kappa^3}{(a^2 + 4\kappa^2G\Lambda_-)\cdot {U_0}^{3/2}}
\label{33}
\end{equation}
where $U = |U(z_0)| > 0$. From (\ref{29}) for the equilibrium radius we have

\begin{equation}
R_0 = \frac{2\kappa z_0}{{U_0}^{1/2}\cdot (a^2 + 4\kappa^2G\Lambda_-)^{1/2}} .
\label{34}
\end{equation}
{}From this formula it is easy to notice that for the de Sitter spaces
$(\Lambda >0)$ the radius is maximal, when \it a \rm is minimal. So to get
the macroscopic bubbles it is worth to set $\Lambda_+ = \Lambda_- = \Lambda$.
In  this case the domain wall separates the  phases  with  the  same  vacuum
energy. Then $a \sim G^2\kappa^4$  which is negligible compared to the second
term in brackets. The charge enters the formulae  (\ref{33}), (\ref{34}) only
through $U_0$ and has very small influence on the values $m_0$ and $R_0$. So
we can set $Q = 0$. With this assumptions one finds  that  \it U  \rm reaches
its maximum $U_0 \sim 1$ at the point $z_0 \sim 1$. This implies that

$$
 m_0 \sim - \frac{\kappa}{G\Lambda} , \\
 R_0 \sim \frac{1}{\sqrt{G\Lambda}} .
$$

Considering a scenario for domain wall production in models with spontaneous
breaking of some gauge symmetry group, it is easy to see that $\Lambda \sim
\alpha^{-1}{M_G}^4$, $\kappa \sim \alpha^{-1}{M_G}^3$ and the thickness of
the wall $d \sim \frac{1}{M_G}$, where $M_G$ is the symmetry breaking scale
and $\alpha \sim 10^{-2}$ is  the coupling constant \cite{BKT87,V85,V91}.

  For the different scales of particle physics we yield:

  a). The  Electro-Weak  scale:  $M_G \sim 10^2GeV$.  Then $\Lambda \sim
10^{10}GeV^4$,
$\kappa \sim 10^8GeV^3$  and $R_0 \sim 10^{14}GeV^{-1}$,
$m_0 \sim -10^{36}GeV$. The  radius  of  such
bubble is about 1 cm, and its negative mass is about million tons.

  b). The scale of family symmetry \cite{C91}: $M_G \sim 10^4 \div 10^{10}GeV$.
Then $\Lambda \sim 10^{18} \div 10^{42}GeV^4$, $\kappa \sim 10^{14} \div
10^{32}GeV^3$  and  $R_0 \sim 10^{10} \div 10^{-2}GeV^{-1} \sim
10^{-4} \div 10^{-16}cm$
, $m_0 \sim - (10^{34} \div 10^{28})GeV \sim - (10^{10} \div 10^4)g$.

  c).  The  scale  of  Grand   Unification:  $M_G \sim 10^{15}GeV$.   Then
$\Lambda \sim 10^{62}GeV^4$, $\kappa \sim 10^{47}GeV^3$ and thus
$R_0 \sim 10^{-12}GeV^{-1}$, $m_0 \sim -10^{23}GeV \sim 0,1g$. This is a very
small radius and  there  arises  the  question
of the validity of the thin-wall approximation ($ d \sim 10^{-15}GeV^{-1}$).

  We see that in all the cases the mass of the bubble is negative and  it
exhibits a strong gravitational repulsion.

\section{The problem of stability.}

  ~~~~Unfortunately this equilibrium state seems to be  unstable.  Potential
(\ref{32}) has the single maximum. In contrary to the  case  with positive
$m$ \cite{BKT87}, the charge does not stabilize the  bubble.  Perhaps one
could find the stable equilibrium states of bubble with negative mass in
some nonvacuum models or in models with rotating bubble.

  In \cite{GKHK78} it is suggested that the inclusion of a nonvanishing
angular momentum might stabilize the shell.  The  surface  energy-momentum
density tensor of a charged rotating shell is given by \cite{I70,L84}

\begin{equation}
T^i_j = - \sigma u^i u_j + \delta^i_j \sigma ,\qquad i,j = 0, 2, 3
\label{36}
\end{equation}
It consists of a mixture of two perfect fluids. The first term can be
interpreted as "dust" particles with negative  energy  density.  The second
one represents a domain wall. The negative mass  "dust" increases the
repulsive character of the bubble and  can  keep  it static. Exterior for of
a spinning charged shell is the  Kerr-Newman metric. Our simple analysis
based on spherical symmetrical equations is not  valid  in  this  case  and
model  (\ref{36})  requires  further investigations.

If the problem of stability will be solved there will be an interesting
possibility of creation of the static singular shell of macroscopic size and
with repulsive gravitational field. Radius and repulsive features of such
objects depend on the scale of symmetry violation and can be varied slowly
by changing the forces stabilizing the bubble.

\bigskip
{\large \bf Acknowledgment . }
\medskip

The research described in this publication was made possible in part by
Grant MXL000 from the International Science Foundation.

\end{document}